\documentclass[sigplan,nonacm]{acmart}
\pdftrailerid{}

\usepackage{amsmath}
\usepackage{booktabs}
\usepackage{graphicx}
\usepackage{overpic}
\usepackage{subcaption}
\usepackage{xspace}
\usepackage{enumitem}
\newcommand{\Sref}[1]{\S~\ref{#1}}

\newcommand{\sysname}{HA-NPU\xspace}
\newsavebox{\sysnamelegendbox}
\newcommand{\sysnamegraphics}[7][]{%
    \begingroup
    \sbox{\sysnamelegendbox}{%
        \normalfont\fontfamily{phv}\bfseries
        \fontsize{#6bp}{#6bp}\selectfont\sysname\unskip}%
    \resizebox{#2}{!}{%
        \begin{overpic}[#1,abs,unit=1sp]{#3}
            \put(#4bp,#5bp){%
                \ifdim\wd\sysnamelegendbox>#7bp\relax
                    \resizebox{#7bp}{!}{\usebox{\sysnamelegendbox}}%
                \else
                    \usebox{\sysnamelegendbox}%
                \fi}
        \end{overpic}}%
    \endgroup
}
\newcommand{\para}[1]{\noindent\textbf{#1}\xspace}

\makeatletter
\renewcommand{\@mkauthors@iii}{%
    \gdef\@currentauthors{}%
    \global\setbox\mktitle@bx=\vbox{%
        \unvbox\mktitle@bx
        \centering
        {\@authorfont
        Yinyuan Zhang\textsuperscript{1}\quad
        Daliang Xu\textsuperscript{2,*}\quad
        Xiaolong Huang\textsuperscript{1}\quad
        Wangsong Yin\textsuperscript{1}\\
        Yun Ma\textsuperscript{1}\quad
        Mengwei Xu\textsuperscript{2}\quad
        Gang Huang\textsuperscript{1}\par}
        \vspace{3pt}
        {\@affiliationfont
        \textsuperscript{1}Peking University\quad
        \textsuperscript{2}Beijing University of Posts and Telecommunications\par}
        \vspace{2pt}
        {\@affiliationfont\ttfamily
        \{yinyuanzhang16,yws\}@stu.pku.edu.cn\quad
        \{huangxiaolong,mayun,hg\}@pku.edu.cn\\
        \{xudaliang,mwx\}@bupt.edu.cn\par}
        \vspace{3pt}
    }%
}
\makeatother

\begin{document}
    \title{Empowering Hybrid Attention Models on NPUs}
    \author{Yinyuan Zhang}
    \affiliation{\institution{Peking University}\city{Beijing}\country{China}}
    \email{yinyuanzhang16@stu.pku.edu.cn}
    \author{Daliang Xu}
    \authornote{Corresponding author.}
    \affiliation{\institution{Beijing University of Posts and Telecommunications}\city{Beijing}\country{China}}
    \email{xudaliang@bupt.edu.cn}
    \author{Xiaolong Huang}
    \affiliation{\institution{Peking University}\city{Beijing}\country{China}}
    \email{huangxiaolong@pku.edu.cn}
    \author{Wangsong Yin}
    \affiliation{\institution{Peking University}\city{Beijing}\country{China}}
    \email{yws@stu.pku.edu.cn}
    \author{Yun Ma}
    \affiliation{\institution{Peking University}\city{Beijing}\country{China}}
    \email{mayun@pku.edu.cn}
    \author{Mengwei Xu}
    \affiliation{\institution{Beijing University of Posts and Telecommunications}\city{Beijing}\country{China}}
    \email{mwx@bupt.edu.cn}
    \author{Gang Huang}
    \affiliation{\institution{Peking University}\city{Beijing}\country{China}}
    \email{hg@pku.edu.cn}
    \renewcommand{\shortauthors}{Yinyuan Zhang et al.}
    \begin{abstract}
    
Hybrid attention models have emerged as a crucial architecture for Large Language Models (LLMs) (e.g., the Qwen3.5 and Kimi series). Their memory and computational efficiency make them highly attractive for on-device inference, forming a promising synergy with edge Neural Processing Units (NPUs). However, naive execution of these hybrid models on edge NPUs fails to deliver these benefits, often bottlenecking the prefill stage due to severe memory-system inefficiencies and architectural mismatches within the linear attention (LA) layers.
We present \sysname, the first system to enable efficient hybrid attention LLM inference on edge NPUs without modifying the underlying algorithms. \sysname enhances execution efficiency by reorganizing the dataflow of the LA components across three levels: (1) At the core level, it partitions workloads by the head dimension and fuses dependent operators, eliminating cross-core global memory accesses; (2) At the operator level, it reorders execution to consume intermediate tensors immediately, drastically minimizing local-buffer pressure; (3) At the tensor level, it employs dataflow-aware layout planning to minimize transformation overhead between matrix and vector processing units. Compared to competitive baselines, \sysname achieves up to 35.95$\times$ LA kernel speedup and 36.14$\times$ energy reduction, delivering up to 2.03$\times$ faster end-to-end request latency.
    The source code will be made publicly available at \url{https://github.com/yinyuanzhang/HA-NPU}.
    \end{abstract}
    \maketitle

    \section{Introduction}
\label{sec:introduction}

Linear attention (LA) is emerging as a crucial building block for open-source large language models (LLMs) by reducing full attention's quadratic complexity to linear and replacing the growing key-value (KV) cache with a fixed-size recurrent state~\cite{katharopoulos2020linear}. Recent LLMs, such as the Qwen3.5 series (0.8B--397B), Qwen3.8-Flash-Next (e.g., 125B), and Kimi K3 (e.g., 2.8T), adopt delta-rule LA mechanisms  such as Gated DeltaNet (GDN)
and Kimi Delta Attention (KDA)  in hybrid architectures~\cite{gated-deltanet,qwen35,qwen38,kimi-linear,kimi-k3}. LA's memory and computational efficiency make it highly attractive for on-edge inference, where compute capability, memory capacity, and energy are stringently constrained. Such deployments typically rely on energy-efficient Neural Processing Units (NPUs)~\cite{song2024powerinfer,llm-npu,wei2025tmac,hao2026mobile-npu}. Together, LA and edge NPUs form a promising algorithm-hardware synergy.

\textbf{Obstacles to LA Inference on Edge NPUs.} Surprisingly, edge NPU execution fails to deliver these expected benefits.
During prompt prefilling, LA is no faster than full attention (FA) and can even bottleneck end-to-end inference. Profiling Qwen3.5-2B on an Ascend 310P edge NPU reveals that, at 4,096 input tokens, a complete LA layer performs $8.7\times$ fewer multiply-accumulate operations than full attention, yet runs $3.02\times$ slower. Furthermore, LA consumes $75.16\%$--$76.99\%$ of the total prefill latency across Qwen3.5 models (Figure~\ref{fig:linear-attention-prefill-share}), with a dismal matrix-unit utilization of $0.20\%$--$0.41\%$. This inefficiency stems from excessive data movement caused by an architectural mismatch between LA execution patterns and edge NPUs. Specifically, we identify three major challenges across the two-level memory hierarchy:

\noindent $\bullet$ \textit{At the global memory (GM) level, fine-grained chunk-head mapping breaks NPU-preferred coarse-grained tensor locality.} LA involves chunk-independent preprocessing and output computation, and sequence-dependent state updating. To maximize GPU multithreading, frameworks like Flash Linear Attention (FLA) map preprocessing and output computation by chunks (without dependency) and state updating by heads (with sequence dependency)~\cite{yang2024fla}. While this mapping boosts parallelism, it triggers frequent cross-core communication that GPUs can hide via massive multithreading. However, NPUs typically process coarse-grained tensor blocks without such mechanisms, resulting in severe data movement overhead where GM accesses account for 56.2\% of LA latency on the Ascend 310P.

\noindent $\bullet$ \textit{At the on-core buffer level, limited cache capacity and conflicting tensor-layout preferences incur additional data movement.}
\begin{itemize}[leftmargin=0pt, itemindent=2em, labelsep=0.5em, nosep]
    \item[(i)] \textit{LA's default execution order overloads local-buffer capacity.} Efficient execution requires keeping a tensor in the cache until all dependent operators consume it. However, LA generates intermediate tensors long before their final use. Because the on-core cache is too small for keeping all these long-lived tensors, the system repeatedly spills and reloads data. As a result, tensor spilling inflates logical GM traffic by $93.4\%$ for GDN and $63.8\%$ for KDA (Figure~\ref{fig:on-chip-capacity-traffic}).
    
    \item[(ii)] \textit{Conflicting tensor-layout preferences between matrix and vector processing units.} LA frequently alternates between matrix (preferring blocked layouts) and vector units (preferring contiguous layouts) for operations like matrix multiplication, scaling, masking, and normalization. While existing systems rely on layout transformations to bridge this gap, they incur substantial overhead~\cite{niu2024smartmem}, consuming 41.58\% of cross-unit execution time and increasing memory traffic by 16.7\% in our profiling. Consequently, \sysname needs to optimally decide whether and where to perform these transformations—an NP-hard problem complicated by cross-chunk state propagation, branched dataflows, and long operator chains.
\end{itemize}

\noindent \textbf{Our approach} This paper presents \sysname, the first system to enable efficient linear-attention LLM inference on edge NPUs without modifying the underlying attention algorithms. We abstract the delta-rule-based LA operators (e.g., GDN and KDA) into a unified Map-Reduce-Map inference paradigm (\Sref{sec:linear-attention-llms}). Building upon this paradigm, our core insight is that reorganizing the LA execution flow to match edge NPU architectural characteristics maximizes data reuse and minimizes data movement while strictly preserving recurrent state dependencies. Specifically: (1) At the core level, we partition workloads by the head dimension and map the entire partition to a single core, avoiding the cross-core dependencies and global memory accesses. (2) At the operator level, we reorder execution so that intermediate tensors are consumed immediately, reducing the cache footprint. (3) At the tensor level, layout transformations are triggered only when they guarantee end-to-end performance gains, thereby minimizing transform overhead. The corresponding novel techniques are detailed as follows:

\textbf{Head-level mapping with cross-operator fusion} (\Sref{sec:recurrence-aware-elastic-fusion}) addresses the frequent GM accesses caused by chunk-head mapping. Our key insight is that since edge NPUs typically have fewer cores than attention heads, head-level parallelism alone saturates the hardware; chunk-level parallelism merely adds severe cross-core GM synchronization overhead. Thus, \sysname partitions workloads by head and fuses dependent operators onto a single core, keeping recurrent states and intermediates in on-core buffers to eliminate cross-core reduction. However, rigid head mapping causes tail-phase load imbalance when the head count is not a multiple of the core count. To resolve this, \sysname proposes sub-head parallelism to the remainder. For instance, mapping a 16-head Qwen model onto a 10-core Ascend 310P leaves 6 tail heads. \sysname splits these remaining heads column-wise along the non-reduction dimension, fully utilizing all 10 cores in the tail phase.

\textbf{Lifetime-aware operator reordering} (\Sref{sec:state-aware-on-chip-graph-reorganization}) mitigates local-buffer pressure from long-lived intermediate tensors. Our key insight is that operators can execute once inputs are ready, safely bypassing LA's default order. \sysname reorders execution to consume intermediates immediately in two scenarios: (i) \textit{Cross-chunk:} To prevent two state versions from coexisting in the cache, \sysname shifts from a \textit{state-first} to a \textit{chunk-first} order, delaying state commits to enable in-place updates. (ii) \textit{Intra-chunk:} Default scheduling interleaves GEMMs and matrix inversions, temporarily idling operands and bloating cache. \sysname adopts a \textit{GEMM-first} order, decoupling inversions and grouping shared-input operations for immediate on-core reuse.

\textbf{Dataflow-aware layout transformation planning} (\Sref{sec:interface-layout-planning}) minimizes layout transformation overhead between matrix and vector units. While optimal layout selection is NP-hard, our key insight is that LA dataflows and NPU constraints offer three structural opportunities to drastically prune the search space: (i) decoupling cross-chunk decisions into isolated intra-chunk subproblems, since only the recurrent state crosses boundaries; (ii) using the strict blocked-layout mandate of matrix units as definitive anchors to simplify branched planning; and (iii) merging contiguous vector-operator chains to evaluate cumulative layout efficiency. Exploiting these properties, \sysname employs dynamic programming to generate an optimal offline layout plan.

\noindent\textbf{Implementation and evaluation.}
\sysname is a general inference system targeting the universal
\textit{matrix-vector} compute paradigm and constrained memory hierarchies
inherent to the architecture of modern edge
NPUs~\cite{ascendc-guide,mahurin2023hexagon,rico2024xdna,intel-npu-library}.
We select Ascend as our representative platform for its open-source stack
and explicit hardware control~\cite{huawei2025cann-open,ascendc-guide},
instantiating the framework atop vLLM-Ascend with
$\sim$13K lines of C/C++.
We comprehensively evaluated \sysname's GDN and KDA kernels on two
representative edge NPUs, Ascend 310P and 310B. For end-to-end inference, we 
evaluated six GDN- and KDA-based LLMs (Qwen3.5-0.8B/2B/4B/9B,
AFM-4.5B, and the MoE-based Ling-3.0-tiny) across diverse text and
multimodal benchmarks. Compared with four competitive baselines---vLLM-Ascend~\cite{kwon2023vllm,vllm-ascend},
FLA~\cite{yang2024fla,gated-deltanet}, FLA + MegaKernel~\cite{hazyresearch2025megakernels,cheng2026mpk}, and FlashQLA~\cite{flashqla2026}---\sysname achieves
$1.69$--$35.95\times$ kernel speedups and $1.76$--$36.14\times$ kernel energy
reductions. End-to-end, it delivers up to $2.03\times$ faster
inference and up to $48.0\%$ lower device energy per request.
Furthermore, compared with FLA, \sysname reduces peak kernel memory by up
to $54.6\%$ and improves GEMM-normalized Matrix-unit utilization by
up to $5.24\times$.

\textbf{Contributions} are summarized as follows:
\begin{itemize}[leftmargin=1.25em,labelsep=0.45em,nosep]
    \item Characterize linear-attention prefilling on edge NPUs and identify a data movement problem spanning frequent GM access, buffer capacity, and tensor layout transformation.
    \item Present \sysname, the first efficient NPU-oriented linear-attention inference system with three novel techniques: head-level mapping, lifetime-aware operator reordering, dataflow-aware layout transformation planning.
    \item  Evaluate on representative GDN and KDA mechanisms, 2 Ascend NPUs, and 6 LLMs, showing consistent kernel, end-to-end latency, energy, and memory improvements.
\end{itemize}

    \section{Background and Motivation}

\subsection{Linear-Attention-based LLMs}
\label{sec:linear-attention-llms}

\para{Rise of Linear-Attention Hybrids.}
Recent frontier open-weight models increasingly adopt hybrid architectures
that replace most dense attention layers with linear attention for better
efficiency. Notably, leading models such as the Qwen series (e.g.,
Qwen3.5, Qwen3.8) adopt GDN~\cite{qwen35,qwen38}, while Kimi K3 and
GLM-5.3-Flash utilize KDA~\cite{kimi-k3,glm53-flash-kda}.
By replacing growing KV caches with fixed-size recurrent states, LA avoids
the quadratic complexity of full attention, drastically reducing the
computational and memory overheads during inference~\cite{gated-deltanet,kimi-linear}.
Crucially, these architectural advantages are particularly indispensable for edge devices, where strict memory and compute constraints make linear attention essential for viable deployments.

\begin{figure}[t]
    \centering
    \captionsetup[subfigure]{
        skip=2pt,
        justification=centering,
        singlelinecheck=false
    }

    \begin{subfigure}[t]{\columnwidth}
        \centering
        \includegraphics[page=27,width=0.96\linewidth]{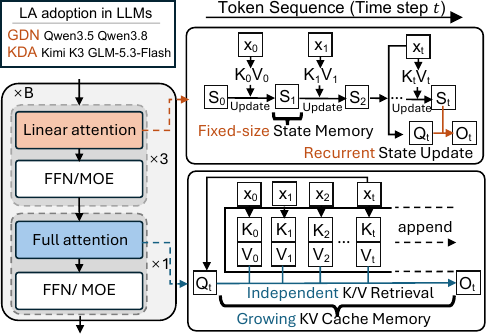}
        \caption{Hybrid linear-attention LLM architecture}
        \label{fig:linear-attention-architecture-hybrid}
    \end{subfigure}\par\vspace{5pt}

    \begin{subfigure}[t]{\columnwidth}
        \centering
        \includegraphics[page=10,width=0.96\linewidth]{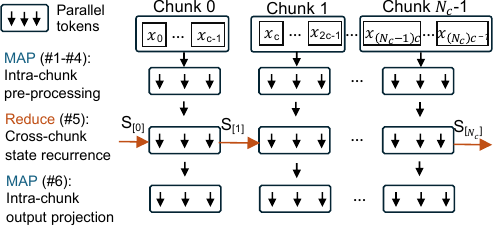}
        \caption{Map--Reduce--Map execution of chunkwise GDN}
        \label{fig:linear-attention-architecture-pipeline}
    \end{subfigure}

    \caption{Architecture and chunkwise execution of the Qwen3.5 (GDN)
    hybrid linear-attention LLM.}

    \label{fig:linear-attention-architecture}
\end{figure}

\para{Architecture of Linear-Attention Hybrids.}
Figure~\ref{fig:linear-attention-architecture}(\subref{fig:linear-attention-architecture-hybrid})
illustrates a typical hybrid architecture that periodically interleaves
linear and full attention blocks (e.g., a 3:1 ratio in Qwen3.5 and
Kimi K3~\cite{qwen35,kimi-k3}). Unlike full attention, which stores a
sequence-growing KV cache, linear attention maintains a fixed-size
recurrent state per head. At step $t$, it updates the previous state
$S_{t-1}$ using $K_t$ and $V_t$, then computes the output with $Q_t$ and
the updated $S_t$. For an input of $T$ tokens, this recurrent formulation
reduces full attention's $\mathcal{O}(T^2)$ compute and $\mathcal{O}(T)$
memory costs to $\mathcal{O}(T)$ and $\mathcal{O}(1)$,
respectively~\cite{katharopoulos2020linear}. However, this state propagation
inherently introduces a strict serial dependency $(S_{t-1} \rightarrow S_t)$,
trading sequence-level parallelism for resource efficiency.

\para{Chunkwise Map-Reduce-Map Execution.}
While token-by-token recurrence supports decoding, to mitigate its severe
serialization bottlenecks during prefilling, modern implementations
partition inputs into chunks, casting execution into a Map--Reduce--Map
paradigm (Figure~\ref{fig:linear-attention-architecture}(\subref{fig:linear-attention-architecture-pipeline})).
Using GDN as an example, an initial \textbf{Map} phase
(operators~\#1--\#4: Gate, KKT, Solve, and W/U) precomputes chunk-local
transformations to capture token-level dependencies.
This enables the subsequent \textbf{Reduce} phase (operator~\#5: H)
to perform state updates via chunk-level matrix operations rather than
token-by-token recurrence, incorporating the incoming state $S_{c-1}$
to yield the updated state $S_c$.
Finally, a subsequent \textbf{Map} phase (operator~\#6: O) combines
these resolved states with local causal interactions to compute the
final token outputs.
Consequently, both Map phases execute independently across chunks,
successfully confining the sequence-dependent bottleneck exclusively
to the Reduce phase.

\subsection{Edge NPUs} 

On-device LLM inference commonly targets NPUs since their specialized
data-parallel engines provide high throughput and energy efficiency under
tight compute, memory, and power budgets~\cite{llm-npu}.

\begin{figure}[t]
    \centering
    \includegraphics[page=3,width=\linewidth]{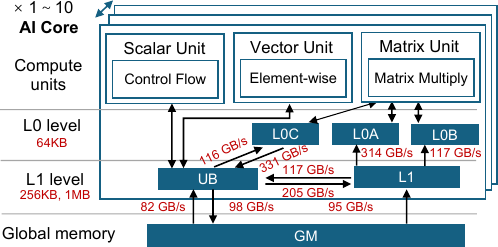}
    \caption{Representative Edge NPU architecture (Ascend 310 series).}
    \label{fig:ascend-npu-architecture}
\end{figure}

\para{NPU Architecture and Memory Subsystem.}
A widely adopted NPU architecture relies on three specialized processing units:
a Scalar unit for control flow, a Vector unit for general data-parallel operations
(e.g., normalization), and a dedicated Matrix unit for compute-intensive tensor
operations (e.g., GEMM) (e.g., Huawei Ascend, Qualcomm Hexagon, AMD XDNA, and Intel
NPUs)~\cite{mahurin2023hexagon,ascendc-guide,rico2024xdna,intel-npu-library}.
To support these processing units, NPUs employ a memory hierarchy comprising a high-capacity
Global Memory (GM) and fast, explicitly managed on-core buffers.
Using Ascend as a representative example (Figure~\ref{fig:ascend-npu-architecture}),
each core features a Unified Buffer (UB) for Vector operations,
L0 buffers for Matrix operations, and L1 for staging reusable
tiles~\cite{zhou2025squeezing,ascendc-guide}.
Crucially, memory bandwidth varies significantly across these data paths; in our
representative profiling on Ascend 310P, the direct on-core forwarding
($\text{UB}\!\rightarrow\!\text{L1}$), GM-to-buffer
($\text{GM}\!\rightarrow\!\text{L1}$), and cross-GM
($\text{UB}\!\rightarrow\!\text{GM}\!\rightarrow\!\text{L1}$)
bandwidths are 205~GB/s, 95~GB/s, and 48~GB/s, respectively.

\para{NPU Programming Model.}
Unlike GPU SIMT (\emph{Single-Instruction, Multiple-Threads}) architectures that rely on dynamic thread scheduling, NPUs expose parallelism primarily through explicit data mapping and SIMD execution~\cite{hao2026mobile-npu,rico2024xdna}. This parallelism operates at two levels. Across AI Cores, the workload is parallelized by mapping distinct data partitions to available cores. Within each AI Core, massive data-level parallelism is achieved through \emph{single-instruction, multiple-data} (SIMD) execution strictly within the Matrix and Vector processing units~\cite{ascendc-guide}.

\begin{figure}[t]
    \centering
    \captionsetup[subfigure]{position=bottom,
        justification=centering,singlelinecheck=false,skip=3pt}
    \begin{subfigure}[t]{0.47\columnwidth}
        \centering
        \includegraphics[width=\linewidth,trim=0bp 0bp 127.22385bp 12bp,clip]{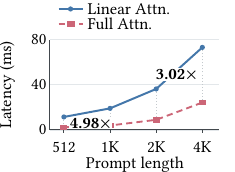}
        \caption{Layer-level latency}
        \label{fig:linear-attention-prefill-layer}
    \end{subfigure}\hfill
    \begin{subfigure}[t]{0.53\columnwidth}
        \centering
        \includegraphics[width=\linewidth,trim=112.82115bp 0bp 0bp 12bp,clip]{figures/linear_attention_prefill_bottleneck.pdf}
        \caption{Model-level latency}
        \label{fig:linear-attention-prefill-model}
    \end{subfigure}
    \caption{Prefill performance on Ascend 310P.}
    \label{fig:linear-attention-prefill-share}
\end{figure}

\subsection{Gaps between Linear Attention and Edge NPUs}
\label{sec:linear-attention-npu-gaps}
Despite the theoretical efficiency of LA and the high throughput of NPUs, we
surprisingly find that naive LA execution severely underperforms during the
prefill stage, often bottlenecking end-to-end inference. Profiling Qwen3.5-2B/4B
on an Ascend 310P shows LA is $3.02$--$4.98\times$ slower than full attention
for 512--4K tokens
(Figure~\ref{fig:linear-attention-prefill-share}(\subref{fig:linear-attention-prefill-layer})),
disproportionately consuming $75.16\%$--$80.12\%$ of the total prefill time
(Figure~\ref{fig:linear-attention-prefill-share}(\subref{fig:linear-attention-prefill-model})).
We then dig into the underlying reasons and find that LA prefill suffers from
a memory-system bottleneck dominated by data movement. We trace this
bottleneck to architectural mismatches across three execution levels: the
core level, the operator level, and the tensor level.

\textbf{Core level: Chunk-head mapping breaks tensor locality.}
To maximize GPU parallelism, existing chunkwise LA executions~\cite{gated-deltanet}
parallelize chunk-independent Map phases while serializing head-dependent
Reduce phases, triggering frequent cross-core communication.
While GPUs hide this GM access overhead via high bandwidth and massive
multithreading, this fine-grained mapping severely clashes with NPU
architectures, which process coarse-grained tensor blocks rather than fine-grained threads.
Consequently, tensors must be repeatedly
flushed to and reloaded from GM, inflating GM accesses to $56.2\%$ of total LA latency for Qwen3.5-2B on Ascend 310P.
 Starved for data, effective Matrix-unit
utilization plummets to a mere $0.20\%$--$0.41\%$---a stark contrast to the
$16.8\%$--$29.0\%$ achieved by full attention
(Table~\ref{tab:fa-la-compute-memory}).

\begin{figure}[t]
    \centering
    \graphicspath{{figures/}}
\captionsetup[subfigure]{skip=2pt,justification=centering,singlelinecheck=false}
\makebox[\linewidth][c]{%
    \includegraphics[page=3,width=5.60cm]{on_chip_capacity_traffic.pdf}}
\par\vspace{2pt}\noindent
\begin{subfigure}[t]{0.485\linewidth}
    \centering
    \includegraphics[page=1,width=\linewidth]{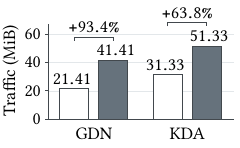}
    \caption{GM traffic}
\end{subfigure}\hfill
\begin{subfigure}[t]{0.485\linewidth}
    \centering
    \includegraphics[page=2,width=\linewidth]{on_chip_capacity_traffic.pdf}
    \caption{L1-to-L0 traffic}
\end{subfigure}
\space
    \caption{Comparison of memory traffic between ideal tensor residency and actual spill/reload on Ascend 310P ($T=2048$, $H=10$, $C=64$).}
    \label{fig:on-chip-capacity-traffic}
\end{figure}

\textbf{Operator level: Default execution order overloads local buffers.}
Within an AI core, compute units heavily rely on limited-capacity local
buffers (e.g., UB for Vector and L0 for Matrix units). 
However, LA's default semantic execution order generates intermediate tensors long before their final consumption. Because these
tensors remain live across multiple intervening operators, their
overlapping lifetimes quickly exceed the strict capacity limits of the
on-core cache~\cite{ahn2020serenity}. Unable to maintain tensor residency, the system triggers
severe capacity-induced thrashing---repeatedly spilling and reloading
intermediate data through the memory hierarchy~\cite{ivanov2021data_movement,dao2022flashattention}. This redundant data
movement imposes substantial memory overhead: our capacity ablation reveals that tensor
spilling inflates logical GM traffic by $93.4\%$ for GDN and $63.8\%$ for KDA
(Figure~\ref{fig:on-chip-capacity-traffic}).

\begin{table}[t]
    \centering
    \caption{Matrix processing unit execution and per-layer history memory of native Full- and
    Linear-Attention cores in Qwen3.5-2B prefill on one Ascend 310P.}
    \label{tab:fa-la-compute-memory}
    \footnotesize
    \setlength{\tabcolsep}{1.5pt}
    \renewcommand{\arraystretch}{1.05}
    \begin{tabular}{@{}rccc@{\hspace{4pt}}ccc@{}}
        \toprule
        & \multicolumn{3}{c}{\textbf{Full Attention}}
        & \multicolumn{3}{c}{\textbf{Linear Attention}} \\
        \cmidrule(lr){2-4}\cmidrule(l){5-7}
        \textbf{Tokens}
        & \shortstack{\textbf{MACs}\\[-1pt]\textbf{(GMAC)}}
        & \shortstack{\textbf{Cube}\\[-1pt]\textbf{util.}}
        & \shortstack{\textbf{K/V cache}\\[-1pt]\textbf{(MiB)}}
        & \shortstack{\textbf{MACs}\\[-1pt]\textbf{(GMAC)}}
        & \shortstack{\textbf{Cube}\\[-1pt]\textbf{util.}}
        & \shortstack{\textbf{State}\\[-1pt]\textbf{(MiB)}} \\
        \midrule
        {\normalsize 512}   & {\normalsize 1.08}  & {\normalsize 16.8\%}
                            & {\normalsize 1.0}   & {\normalsize 0.75}
                            & {\normalsize 0.20\%} & {\normalsize 1.0} \\
        {\normalsize 1,024} & {\normalsize 4.33}  & {\normalsize 22.0\%}
                            & {\normalsize 2.0}   & {\normalsize 1.51}
                            & {\normalsize 0.29\%} & {\normalsize 1.0} \\
        {\normalsize 2,048} & {\normalsize 17.31} & {\normalsize 26.7\%}
                            & {\normalsize 4.0}   & {\normalsize 3.02}
                            & {\normalsize 0.37\%} & {\normalsize 1.0} \\
        {\normalsize 4,096} & {\normalsize 69.25} & {\normalsize 29.0\%}
                            & {\normalsize 8.0}   & {\normalsize 6.04}
                            & {\normalsize 0.41\%} & {\normalsize 1.0} \\
        \bottomrule
    \end{tabular}
\end{table}

\begin{figure}[t]
    \centering
    \captionsetup{skip=3pt}
    \captionsetup[subfigure]{position=bottom,
        justification=centering,singlelinecheck=false,skip=2pt}
    \begin{subfigure}[t]{0.31\columnwidth}
        \centering
        \vspace{0pt}
        \includegraphics[page=17,width=\linewidth]{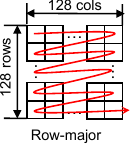}\par
        \vspace{0.1272\linewidth}
        \caption{Dense layout}
        \label{fig:ascend-matrix-layouts-dense}
    \end{subfigure}\hfill
    \begin{subfigure}[t]{0.675\columnwidth}
        \centering
        \vspace{0pt}
        \includegraphics[page=18,width=\linewidth]{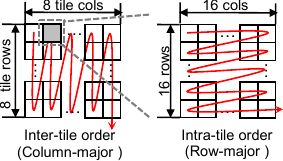}
        \caption{Blocked layout}
        \label{fig:ascend-matrix-layouts-blocked}
    \end{subfigure}
    \caption{Dense and Blocked memory layouts of a $128\times128$ FP16 matrix. Red arrows show element order in memory.}
    \label{fig:ascend-matrix-layouts}
\end{figure}

\textbf{Tensor level: Conflicting layout preferences inflate transformation overhead.}
Within an AI core, the LA dataflow frequently alternates between Matrix
processing units (e.g., for GEMMs) and Vector processing units (e.g., for
normalization). However, these units impose fundamentally
incompatible memory layouts: Matrix processing units mandate two-level
blocked layouts to align with tiled multipliers, whereas Vector processing
units strictly favor contiguous dense layouts for regular sequential
access (Figure~\ref{fig:ascend-matrix-layouts}). Consequently, as tensors are
continuously handed off between units, their physical layouts must be
repeatedly transformed even when logical values remain
unchanged~\cite{niu2024smartmem}. LA's highly interleaved operator chain
severely amplifies this structural conflict, turning logical tensor handoffs into pure data-movement overhead. 
In our profiling of Qwen3.5-2B, these
redundant layout transformations consume $41.58\%$ of the cross-unit execution
time and increase memory traffic by $16.7\%$.

\section{Design Overview}
\label{sec:design-overview}

\begin{figure}[t]
    \centering
    \includegraphics[page=24,width=\columnwidth]{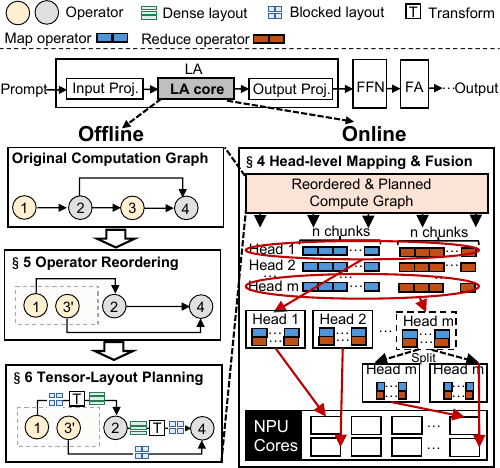}
    \caption{Overview of \sysname.}
    \label{fig:design-overview}
\end{figure}

\para{Design Goal}
\sysname is a general, memory-efficient LA inference system designed specifically for resource-constrained edge NPUs to significantly reduce prefill latency and energy consumption. 
It operates as a transparent, drop-in engine that integrates seamlessly into existing frameworks (e.g., vLLM-Ascend) without altering original attention semantics. It natively supports SOTA delta-rule architectures (e.g., GDN, KDA) and easily extends to others (e.g., RetNet, GLA).

\para{Workflow.}
To alleviate the severe memory-system bottlenecks of LA
prefilling on edge NPUs, \sysname shifts the execution paradigm from
sequence-dimension parallelism to operator-dataflow locality. Abstracting
the workload into a chunkwise Map--Reduce--Map paradigm, it systematically
reconstructs the inference pipeline across three architectural levels:
(1) \textit{At core level}: It partitions workloads by head and fuses dependent operators onto a single core, eliminating cross-core GM accesses.
(2) \textit{At operator level}: It reorganizes operator execution order to consume intermediate tensors immediately, mitigating spill and reload traffic within limited on-core buffers.
(3) \textit{At tensor level}: It coordinates tensor layouts to resolve format conflicts between matrix and vector processing units, minimizing transformation overheads.

\begin{itemize}[leftmargin=0.75em,labelsep=0.25em,nosep]
\item \textbf{Offline Stage (Operator \& Tensor Levels)}. Given an LA
computation graph (Figure~\ref{fig:design-overview}), \sysname generates a deterministic static execution plan, 
\textit{reordered \& planned graph}. It sequentially applies lifetime-aware operator
reordering (\Sref{sec:state-aware-on-chip-graph-reorganization}) to determine
a memory-efficient execution sequence, and dataflow-aware layout planning
(\Sref{sec:interface-layout-planning}) to assign exact physical formats and
transformation sites for all tensors.

\item \textbf{Online Stage (Core Level)}
Taking the input projection's $Q/K/V$ tensors, gates, and initial state, \sysname executes this \textit{reordered and planned graph} via head-level mapping with fusion (\Sref{sec:recurrence-aware-elastic-fusion}). This mapping pins each head to a dedicated AI Core and elastically splits the execution tail,
maximizing hardware saturation and preserving on-core locality, yielding the final attention outputs and updated states for downstream layers and subsequent decoding.
\end{itemize}

\section{Head-Level Mapping with Cross-Operator Fusion}
\label{sec:recurrence-aware-elastic-fusion}

In chunkwise LA, the overall computation is organized across independent
attention heads ($H_1, \dots, H_m$) and sequence chunks ($C_1, \dots, C_n$).
The execution structurally alternates between Map and Reduce operators
with fundamentally conflicting execution patterns: Map operators
allow chunk-level concurrent execution, whereas the Reduce operator
enforces strict head-level sequential execution.

\begin{figure*}[t]
    \centering
    \includegraphics[page=2,width=0.98\textwidth]{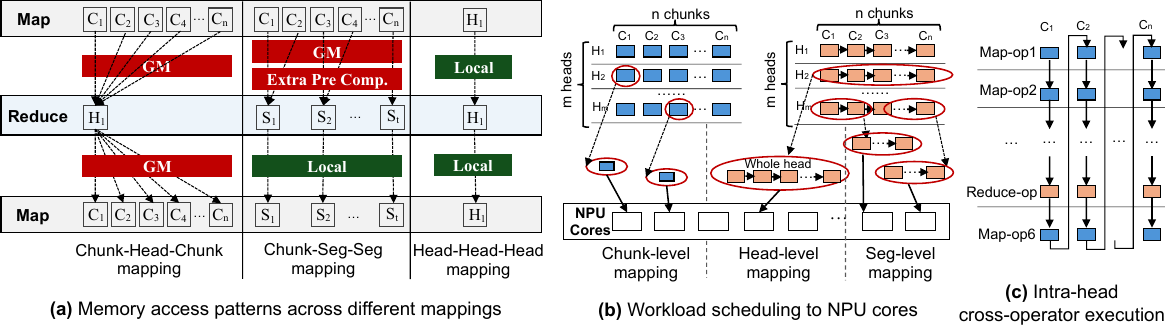}
    \caption{Illustration of Chunk-Head-Chunk, Chunk-Seg-Seg, and Head-Head-Head execution mappings of
    chunkwise LA.}
    \label{fig:recurrence-aware-elastic-fusion}
\end{figure*}

To execute this interleaved workload, mainstream frameworks
pursue maximum parallelism for massive GPU concurrency, assembling
these granularities into two predominant schemas
(Figure~\ref{fig:recurrence-aware-elastic-fusion}(a,b)):

\begin{itemize}[leftmargin=0.75em,labelsep=0.25em,nosep]
\item \textbf{Chunk-Head-Chunk Mapping:}~\cite{kwon2023vllm,yang2024fla} 
This schema completely decouples the operators to isolate the serial
bottleneck. It executes both Map phases as parallel chunk-level workloads
(e.g., $C_1, \dots, C_n$), interrupted by a strictly sequential head-level
Reduce workload (e.g., updating the recurrent state across
$C_1 \rightarrow C_2 \rightarrow \dots \rightarrow C_n$ within $H_1$).

\item \textbf{Chunk-Seg-Seg Mapping:}~\cite{flashqla2026}
In contrast, this schema pursues concurrency by parallelizing the
bottleneck itself. It executes the first Map phase as parallel chunk-level
workloads (e.g., $C_1, \dots, C_n$), followed by extra pre-computation that
enables concurrent segment-level execution for Reduce and final Map
phases (e.g., splitting $H_1$'s chain into independent segments $S_1(C_1 \rightarrow C_2)$ and $S_2(C_3 \rightarrow C_4)$).

\end{itemize}

\para{Fractured Locality and the Memory Wall.} Fundamentally, both mappings reflect a GPU-centric design philosophy
that prioritizes massive thread concurrency---a strict necessity to hide
memory latency and saturate massive GPU architectures (e.g., an NVIDIA
RTX 4090 requires thousands of concurrent workloads to fully utilize its
128 SMs and 48 warps/SM). However, on edge NPUs, this execution model
fractures data locality and creates a severe memory wall. Because edge
NPUs are strictly bottlenecked by constrained GM bandwidth, independently
scheduling workloads of varying granularities severs on-core data
dependencies, triggering frequent cross-core synchronization that forces
intermediate tensors and recurrent states to be repeatedly materialized
to GM. This massive GM traffic drastically inflates data movement
overhead, severely starving the high-throughput Matrix units
(\Sref{sec:linear-attention-npu-gaps}).

To shatter this memory wall, \sysname shifts the execution paradigm from \textit{pursuing parallelism along the sequence dimension} to \textit{maximizing data locality along the operator dataflow}. 
This shift is motivated by two key insights:

\para{Insight 1: Head-Level Independence Across Operators.}
Despite intra-head serialization during state updates, zero data dependencies exist across heads throughout all Map and Reduce operators, eliminating cross-head synchronization.  

\para{Insight 2: Redundant LA Parallelism on Edge NPUs.} Edge NPUs integrate very few compute units (e.g., 1--10 in Ascend 310, 1 in Qualcomm NPUs)~\cite{ascend310b-spec,ascend310p-spec,hao2026mobile-npu}, relative to the abundant attention heads in modern LLMs (e.g., 16--64 in Qwen3.5)~\cite{qwen35-08b,qwen35-4b,qwen35-model-card}.
Thus, when enough heads are available to occupy all cores, exposing fine-grained chunk-level workloads offers no additional benefit to core-level concurrency.

Leveraging these properties, \sysname adopts a head-centric execution paradigm that guarantees strict on-core tensor reuse while ensuring full NPU saturation. Specifically, it introduces two core mechanisms: (1) \textit{head-level mapping with cross-operator fusion}, which vertically integrates a head's heterogeneous operators into a single workload pinned to a dedicated core; and (2) \textit{computation splitting}, which adaptively partitions tail-end execution to expose additional concurrent workloads when the remaining heads underfill the NPU.

\para{Head-Level Mapping with Cross-Operator Fusion.}
To resolve the granularity mismatch, \sysname adopts the attention head as
its fundamental scheduling unit, mapping independent heads to different
dedicated cores for concurrent execution. Within each core, it fuses the
entire Map--Reduce--Map pipeline into a single kernel. As illustrated in
Figure~\ref{fig:recurrence-aware-elastic-fusion}(c), a core sequentially
processes chunks ($C_1 \rightarrow \cdots \rightarrow C_n$), executing all
operators along the dataflow (Map-op1 $\rightarrow \cdots \rightarrow$
Reduce-op $\rightarrow$ Map-op6) within each chunk $C_i$. This localized
execution confines intermediate tensors and state $S_i$ within on-core
buffers, bypassing the GM memory wall.

\begin{figure}[t]
    \centering
    \captionsetup[subfigure]{position=bottom,justification=centering,singlelinecheck=false,skip=3pt}
    \begin{subfigure}{\linewidth}
        \centering
        \includegraphics[page=15,width=0.97\linewidth]{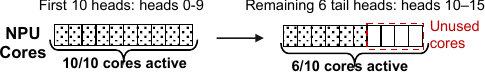}
        \caption{Core underutilization at the execution tail}
        \label{fig:wave-underfill}
    \end{subfigure}
    \par\medskip
    \begin{subfigure}{\linewidth}
        \centering
        \includegraphics[page=16,width=0.97\linewidth]{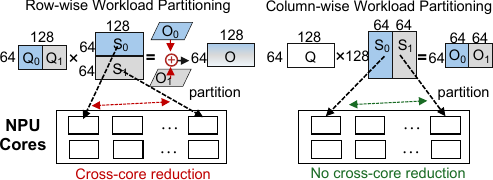}
        \caption{Row versus column splitting of $S$}
        \label{fig:state-splitting}
    \end{subfigure}
    \caption{Comparison of workload partitioning strategies.}
    \label{fig:wave-underfill-and-task-splitting}
\end{figure}

\begin{figure}[t]
    \centering
    \includegraphics[page=3,width=0.98\columnwidth]{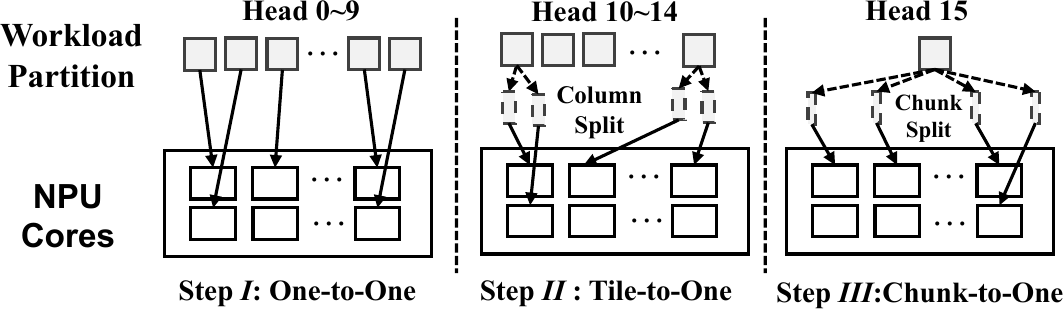}
    \caption{Workflow of \sysname's three-tiered scheduling.}
    \label{fig:intra-request-elastic-fusion}
\end{figure}

\para{Computation Splitting.}
While head-level mapping maximizes data locality, it risks core
underutilization at the execution tail (e.g., mapping 16 heads onto 10 cores
leaves 6 residual heads,
Figure~\ref{fig:wave-underfill-and-task-splitting}(\subref{fig:wave-underfill})).
To eliminate this, \sysname dynamically partitions these residual
heads to expose additional concurrent workloads. A row-wise split of the
recurrent state $S$ mandates costly cross-core reductions. Instead,
\sysname exploits the structural independence of state updates by employing
column-wise partitioning
(Figure~\ref{fig:wave-underfill-and-task-splitting}(\subref{fig:state-splitting})).
Assigning each column block to a distinct core inherently doubles the
concurrent workloads for each selected head, preserving on-core locality
without requiring cross-core reduction.

\para{Putting It Together.}
To preserve on-core tensor reuse and sufficient parallelism, \sysname orchestrates these mechanisms through a three-tiered scheduling pipeline (Figure~\ref{fig:intra-request-elastic-fusion}):
\begin{itemize}[leftmargin=0.75em,labelsep=0.25em,nosep]
\item \textit{Step I (Sufficient Head Parallelism):}
Each head executes on one AI Core, preserving on-core tensor reuse across operators and chunks.
\item \textit{Step II (Underfilled Tails):}
Partitions residual heads along the columns of $S$ to expose independent workloads.
\item \textit{Step III (Extreme Tails):}
Reverts to chunk-head-chunk mapping exclusively for boundary cases, injecting sequence-dimension concurrency to guarantee core saturation.
\end{itemize}

    \section{Lifetime-Aware Operator Reordering}
\label{sec:state-aware-on-chip-graph-reorganization}

While head-resident mapping (\Sref{sec:recurrence-aware-elastic-fusion})
localizes data, sustaining it requires buffering recurrent states,
intermediate tensors, and I/O tensors with overlapping lifetimes. The
default semantic execution order severely exacerbates this pressure by
generating intermediate tensors long before their consumption,
unnecessarily prolonging their lifetimes. For instance, adding a 64-KiB
state to the 244-KiB kernel footprint pushes peak memory to 308~KiB,
exceeding Ascend 310P's 256-KB buffer. This capacity deficit triggers
inevitable cache thrashing, forcing the NPU to repeatedly evict and reload
still-needed tensors, negating on-core locality.

\noindent \textbf{Lifetime-Aware Reordering.}
To eliminate this, our core insight is that operators with ready
inputs are not bound by default semantic sequence. Inspired
by out-of-order execution in modern processors, \sysname reorders operator execution to
prioritize computations that consume intermediate tensors immediately
after generation. This \textit{tensor lifetime minimization} strategy forces the
early release of memory footprints, keeping peak memory within
local buffer capacities. Specifically, \sysname applies this operator
reordering to two critical scenarios: (i) across chunks for recurrent
state updates, and (ii) within a chunk for shared-input GEMMs.

\begin{figure}[t]
    \centering
    \begin{subfigure}{\columnwidth}
        \centering
        \includegraphics[width=0.95\linewidth]{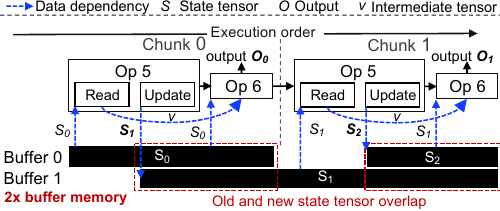}
        \caption{State-first: state update before output}
        \label{fig:state-first-order}
    \end{subfigure}
    \par\medskip
    \begin{subfigure}{\columnwidth}
        \centering
        \includegraphics[width=0.95\linewidth]{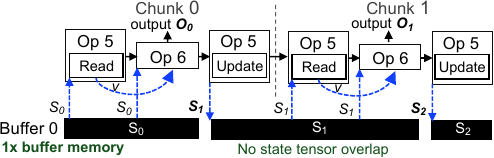}
        \caption{Chunk-first: output before state update}
        \label{fig:chunk-first-order}
    \end{subfigure}
    \caption{Operator (Op) decoupling and reordering for in-place state updates.}
    \label{fig:interface-layout-coupling}
\end{figure}

\noindent $\bullet$ \textit{Cross-Chunk: State-First to Chunk-First Order via Operator Splitting.}
The default \textit{state-first} order computes the new state $S_{c+1}$
(Operator 5) before the chunk output (Operator 6) finishes consuming the
old state $S_c$
(Figure~\ref{fig:interface-layout-coupling}(\subref{fig:state-first-order})).
This lifetime overlap forces $S_c$ and $S_{c+1}$ to co-reside in the on-core
buffer, doubling the state memory footprint. To eliminate this overhead,
\sysname exploits a critical \textit{partial dependency}: while the default
semantics treat Operator 5 as a monolithic step, Operator 6 actually
requires only the incoming state $S_c$ and an intermediate tensor $v_c$,
completely bypassing the need for the final updated state $S_{c+1}$.
\sysname thus decouples Operator 5 and transitions to a \textit{chunk-first} order
(\textit{Operator 5-read} $\rightarrow$ Operator 6 $\rightarrow$
\textit{Operator 5-update},
Figure~\ref{fig:interface-layout-coupling}(\subref{fig:chunk-first-order})),
deferring the state update until $S_c$ is read for the last time.
This enables an in-place overwrite, halving the state memory requirement
while strictly preserving algorithmic recurrence.

\begin{figure}[t]
    \centering
    \begin{subfigure}{\columnwidth}
        \centering
        \includegraphics[page=12,width=0.96\linewidth]{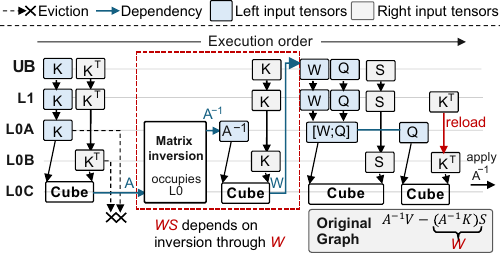}
        \caption{Inversion-first order}
        \label{fig:shared-operand-native}
    \end{subfigure}
    \par\medskip
    \begin{subfigure}{\columnwidth}
        \centering
        \includegraphics[page=13,width=0.96\linewidth]{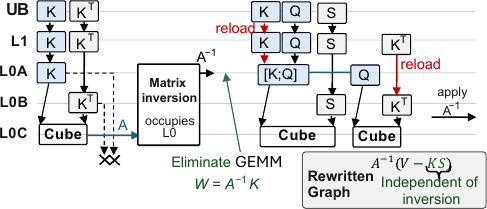}
        \caption{GEMM-first order --- Step I: Remove inversion dependency}
        \label{fig:shared-operand-rewritten}
    \end{subfigure}
    \par\medskip
    \begin{subfigure}{\columnwidth}
        \centering
        \includegraphics[page=14,width=0.96\linewidth]{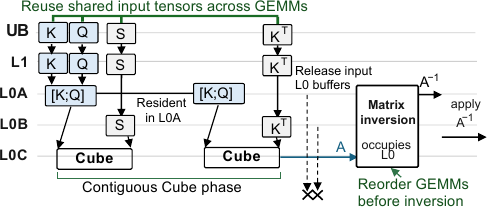}
        \caption{GEMM-first order --- Step II: Reorder for input reuse}
        \label{fig:shared-operand-reordered}
    \end{subfigure}
    \caption{Operator rewriting and reordering enable GEMM-first execution for on-core tensor reuse.}
    \label{fig:shared-operand-analyze}
\end{figure}

\noindent $\bullet$ \textit{Intra-Chunk: Inversion-First to GEMM-First Order via Operator Rewriting.} 
The default \textit{inversion-first order} schedules the matrix inversion ($A^{-1}$) between GEMMs that share input tensors (e.g., $K$, $Q$, $S$, and $K^{\mathsf T}$) (Figure~\ref{fig:shared-operand-analyze}(\subref{fig:shared-operand-native})).
Thus, these shared inputs (e.g., $K^T$) remain temporarily idle across the inversion phase. 
To sustain efficient pipelining, the inversion exhausts L0 buffer with its intermediates, forcing redundant evictions and L1 reloads of idle inputs for subsequent GEMMs.

\begin{figure*}[t]
  \centering
  \includegraphics[page=1,width=0.98\textwidth,trim=0bp 18.5bp 32bp 0bp,clip]{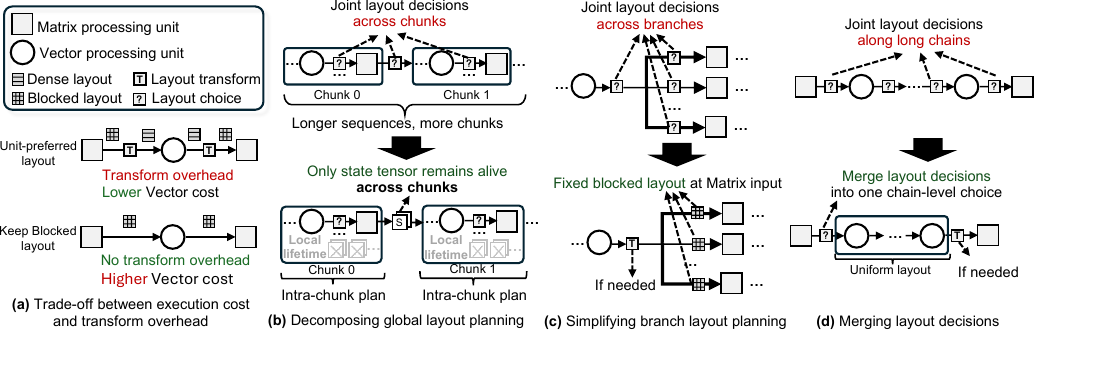}
  \caption{Illustration of dataflow-aware layout planning.}
  \label{fig:layout-planning-coupling}
\end{figure*}

\sysname eradicates these transfers through a methodical two-step
optimization. First, it breaks the strict data dependency of $W = A^{-1}K$
via operator rewriting. By factoring out $A^{-1}$
(Figure~\ref{fig:shared-operand-analyze}(\subref{fig:shared-operand-rewritten})),
the newly formulated GEMM operation directly consumes $K$,
effectively decoupling this computation from the costly matrix inversion.
Second, with this dependency
resolved, \sysname transitions to a \textit{GEMM-first order} by grouping
all shared-input GEMM operators ($KS$, $QS$, $KK^T$, $QK^T$) before the
inversion
(Figure~\ref{fig:shared-operand-analyze}(\subref{fig:shared-operand-reordered})).
Executing them in a contiguous phase maintains these inputs within the L0 buffer for direct reuse, subsequently releasing them to free workspace for the inversion.

\section{Dataflow-Aware Layout Transformation Planning}
\label{sec:interface-layout-planning}

LA frequently alternates between Matrix and Vector computations, triggering
a fundamental layout conflict: Matrix processing units achieve high
throughput with Blocked layouts, whereas Vector processing units favor
contiguous Dense layouts (\Sref{sec:linear-attention-npu-gaps}). This
conflict introduces a strict trade-off
(Figure~\ref{fig:layout-planning-coupling}(a)): transforming layouts incurs
transformation overheads~\cite{niu2024smartmem}, while computing with
mismatched formats suffers execution penalties---a balance dependent on
specific operator types and tensor shapes.
Consequently, \sysname formulates layout planning as a discrete cost
minimization problem to find a plan $\pi^*$ that
minimizes total inference latency:
$\pi^* = \arg\min_{\pi} [T_{exec}(\pi) + T_{transform}(\pi)]$,
where $T_{exec}$ and $T_{transform}$ represent the layout-specific execution
costs and transformation overheads, respectively.

\para{Combinatorial Layout Search.}
Finding the global optimum $\pi^*$ is non-trivial because layout
decisions are deeply coupled across the execution
dataflow~\cite{liu2019neocpu}. Specifically, these dependencies necessitate
joint layout optimization across three structural dimensions:
(i) sequence-level cross-chunk propagation
(Figure~\ref{fig:layout-planning-coupling}(b)),
(ii) shared tensors in multi-branch dataflows
(Figure~\ref{fig:layout-planning-coupling}(c)), and
(iii) cumulative overheads along long Vector chains
(Figure~\ref{fig:layout-planning-coupling}(d)).
Together, these deep couplings formulate the layout optimization as a Partitioned Boolean Quadratic Programming (PBQP) problem. General PBQP is NP-hard~\cite{niu2024smartmem}. For an input partitioned into $N_c$ chunks with up to $m$ layout choices per chunk, the search space explodes to $2^{m * N_c}$. As $N_c$ scales linearly with the input sequence length, naive enumeration becomes intractable for long-context inference.

\para{Pruning the search space.}
Fortunately, the regular LA dataflow and NPU hardware constraints provide three key opportunities to drastically prune the search space:  

\begin{itemize}[leftmargin=0.75em,labelsep=0.25em,nosep]
\item \textit{Decoupling global planning into intra-chunk subproblems.}

Because layout costs propagate strictly via tensor dependencies and only the recurrent state crosses chunk boundaries, intra-chunk layout decisions remain perfectly isolated (Figure~\ref{fig:layout-planning-coupling}(b)). \sysname exploits this by retaining only the minimal intra-chunk cost for limited state-layout combinations, effectively breaking the cross-chunk coupling.

\item \textit{Simplifying branch layout planning through fixed Matrix interfaces.}
We leverage a consistent dataflow characteristic of LA: branched shared
tensors invariably serve as operands for downstream Matrix computation units. Since
these units strictly mandate Blocked inputs, \sysname uses this
terminal constraint as a definitive anchor, directly pruning the
combinatorial branch search into a deterministic, one-time transformation
evaluation (Figure~\ref{fig:layout-planning-coupling}(c)).

\item \textit{Merging Vector chains into a single logical unit.}
For shape-preserving vector chains here, layout transform
costs are independent of placement under our cost model. Since each vector
operator executes no slower in Dense layout, an all-Dense chain with boundary transforms is no more expensive than any mixed-layout assignment.
\sysname therefore evaluates only two candidates for each chain: all-Dense
with boundary transforms, or all-Blocked
(Figure~\ref{fig:layout-planning-coupling}(d)).

\end{itemize}

\para{Finding the optimal layout plan.}
With the search space rigorously bounded, the global optimization reduces to a straightforward dynamic programming (DP) recurrence: $D_i(s) = \min_p [D_{i-1}(p) + c_i(p, s)]$, where $D_i(s)$ represents the minimum accumulated cost up to step $i$ under layout state $s$, and $c_i(p,s)$ is the transition cost from predecessor state $p$.
\sysname evaluates this recurrence in a strict topological order. Crucially, 
at chunk boundaries, the layout state $s$ collapses to retain only the recurrent-state layout, ensuring the search complexity scales linearly ($O(N_c)$) rather than exponentially. 

\para{Global Plan Generation.}
During the DP forward pass, \sysname records the optimal predecessor state for each layout transition. 
The planner then backtracks to assign exact physical layouts to all tensors and pinpoint transformation insertion sites. 
This offline generation yields a static, deterministic execution plan, completely eliminating runtime layout-scheduling overhead.

    \section{Implementation and Evaluation}

\sysname is a general inference system designed around the ubiquitous traits
of modern edge NPUs: utilizing head-level mapping for limited compute
cores, operator reordering for constrained memory hierarchies, and layout
planning for the matrix-vector compute
paradigm~\cite{ascendc-guide,mahurin2023hexagon,rico2024xdna,intel-npu-library}.
We instantiate our prototype on the Ascend 310 series specifically
because its open-source, programmable stack affords the explicit hardware
control necessary for these low-level
optimizations~\cite{ascendc-guide,huawei2025cann-open}.
As a $\sim$13K-line C/C++ kernel-level drop-in replacement, \sysname uses a
\texttt{torch\_npu} binding~\cite{torch-npu} to integrate natively into
vLLM-Ascend~\cite{kwon2023vllm,vllm-ascend} without modifying the high-level
serving pipeline.
Our implementation further addresses two practical concerns:
(1) \textit{Full-Layer Implementation:} An LA layer comprises input/output projections and the LA core. We accelerate these projections via operator fusion and convolution-metadata reuse ($1.62\times$ speedup over native vLLM-Ascend), equipping all baselines with this identical backend to strictly isolate our core improvements during end-to-end comparisons.
(2) \textit{Minimized I-cache pressure:} \sysname shares fixed-shape Cube routines across GEMMs, eliminating redundant scalar control code to reduce I-cache misses.

    \label{sec:evaluation}

\begin{figure*}[t]
    \centering
    \captionsetup[subfigure]{skip=2pt}
    \begin{subfigure}[t]{\textwidth}
        \centering
        \sysnamegraphics{\linewidth}{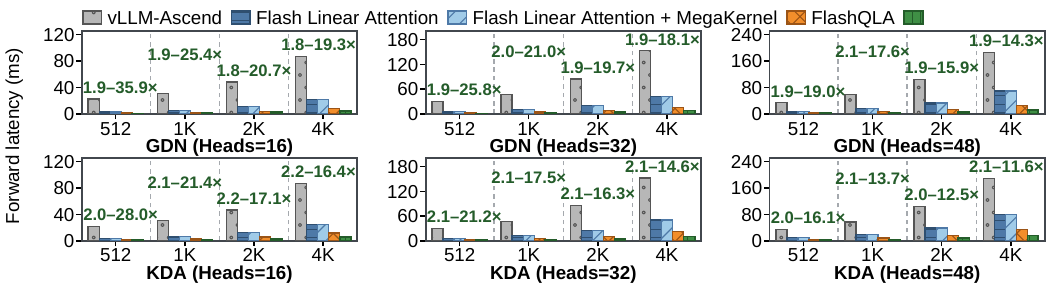}{445.88501}{126.74759}{9}{56.11499}
        \caption{\textbf{Ascend 310P} (X-axis: Sequence length $T$ (tokens))}
        \label{fig:linear-attention-core-performance-310p}
    \end{subfigure}
    \par\vspace{2pt}
    \begin{subfigure}[t]{\textwidth}
        \centering
        \includegraphics[width=\linewidth,trim=0 0 0 13pt,clip]{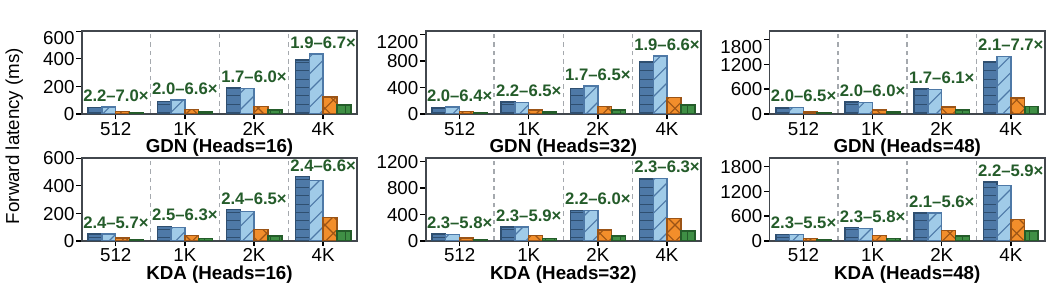}
        \caption{\textbf{Ascend 310B} (X-axis: Sequence length $T$ (tokens); \textbf{vLLM-Ascend is unsupported on 310B}) }
        \label{fig:linear-attention-core-performance-310b}
    \end{subfigure}
    \caption{Forward latency of GDN and KDA under different sequence lengths and head counts
    on Ascend 310P and 310B. 
    Green annotations show the range of speedups achieved by \sysname over the baselines for each configuration.}
    \label{fig:linear-attention-core-performance}
\end{figure*}

\begin{figure*}[t]
    \centering
    \sysnamegraphics[trim=0 5pt 0 0,clip]{\textwidth}{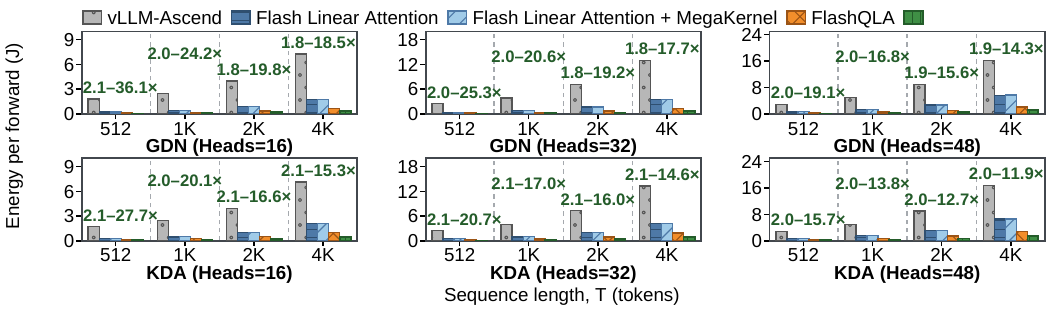}{445.88501}{135.06075}{9}{56.11499}
    \caption{Energy consumption of GDN and KDA under different sequence lengths and head
    counts on Ascend 310P.}
    \label{fig:linear-attention-core-energy}
\end{figure*}

\subsection{Experiment Setups}

\para{Hardware Setup.}
We evaluate \sysname on two edge Ascend NPUs:
Ascend 310P (10 AI Cores, 96 GB LPDDR4X, 8-core CPU)
and Ascend 310B (one AI Core, 24 GB LPDDR4X, 4-core CPU)~\cite{ascend310p-spec,ascend310b-spec,orangepi-aipro20t}.
All experiments use the CANN toolkit~\cite{ascendc-guide} within vLLM-Ascend~\cite{vllm-ascend} for end-to-end evaluation.

\para{Models and Datasets.}
We evaluate two SOTA linear attention mechanisms—GDN~\cite{gated-deltanet} and KDA~\cite{kimi-linear}—using four GDN-based \textit{multimodal models} (Qwen3.5-0.8B/2B/4B/9B)~\cite{qwen35,qwen35-models} and two KDA-based \textit{text-only models} (AFM-4.5B~\cite{afm45-kda-nope} and Ling-3.0-tiny~\cite{ling3-tiny}).
For end-to-end evaluation, we use text QA (TriviaQA~\cite{joshi2017triviaqa}, HotpotQA~\cite{yang2018hotpotqa}) across all models, MMMU~\cite{yue2024mmmu} for GDN-based multimodal evaluation, and MMLU-Pro~\cite{wang2024mmlupro} for text-only KDA models.

\para{Baselines.}
We compare \sysname with four SOTA baselines:
(1) \textit{vLLM-Ascend}~\cite{kwon2023vllm,vllm-ascend} (v0.19.1rc1), the native inference backend;
(2) \textit{Flash Linear Attention (FLA)}~\cite{yang2024fla,gated-deltanet}, using chunkwise execution;
(3) \textit{FLA + MegaKernel}~\cite{hazyresearch2025megakernels,cheng2026mpk}, a single-kernel variant assessing kernel-launch overheads; and
(4) \textit{FlashQLA}~\cite{flashqla2026}, using its fused, segment-parallel execution.
As official FLA and FlashQLA releases lack NPU support, we rigorously ported
their computation flows to Ascend 310P/310B, strictly preserving their principal
fusion and segmentation strategies.
Because official FlashQLA release supports only GDN, we accurately extended its algorithmic logic to a KDA version.
For a controlled comparison, all methods share identical projection and
decoding, a chunk size of 64, and mixed-precision dataflows
(FP16 Q/K/V and FP32 gates/states).

\para{Metrics.} We report core forward latency, end-to-end request latency, peak memory, energy (integrated from 100-ms power samples over a continuous 10-second execution via Ascend's API, divided by total calls), memory traffic (aggregated via hardware counters), and Matrix utilization (hardware-recorded MACs divided by overhead-free device-event timings, normalized to the device's measured peak FP16 GEMM throughput). All experiments report the average of 3–5 runs.

\subsection{Linear-Attention Kernel Performance}

\begin{figure*}[t]
    \centering
    \sysnamegraphics[trim=0 10pt 0 0,clip]{\textwidth}{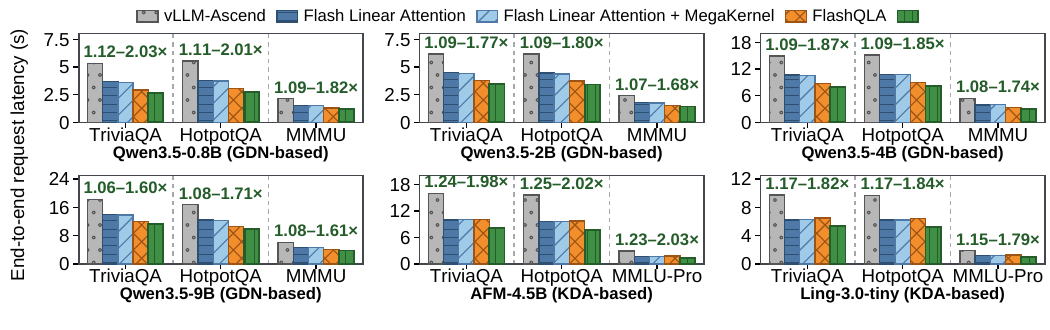}{443.63843}{137.90148}{8}{58.36157}
    \includegraphics[width=\textwidth,trim=0 10pt 0 12pt,clip]{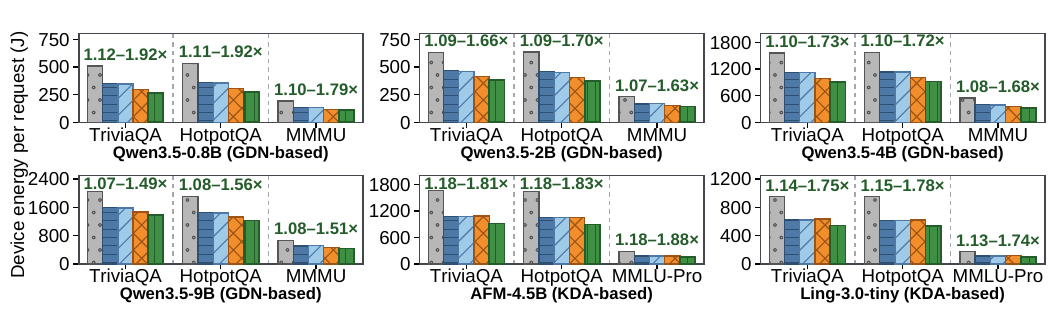}
    \caption{End-to-end request latency and measured device energy per request
    on Ascend 310P.}
    \label{fig:e2e-dataset-latency}
\end{figure*}

We evaluate the core performance of \sysname across a comprehensive sweep of attention formulations, head counts, sequence lengths, and devices. As shown in Figures~\ref{fig:linear-attention-core-performance} and~\ref{fig:linear-attention-core-energy}, \textbf{\sysname consistently delivers superior execution efficiency and energy savings across all configurations.}

\para{Forward Latency.}
\begingroup
\setlength{\emergencystretch}{1em}
\textit{Compared with vLLM-Ascend:} \sysname achieves an
$11.60$--$19.28\times$ speedup on Ascend 310P (4,096 tokens) by replacing
fragmented general-purpose \texttt{torch\_npu} operators with highly fused,
NPU-aligned kernels.
\textit{Compared with FLA:} \sysname achieves a $4.74$--$7.02\times$
speedup (4,096 tokens) by replacing FLA's GM-bound, stage-by-stage execution
with head-level on-core reuse.
\textit{Compared with FLA-MegaKernel:} \sysname maintains a
$4.74$--$7.69\times$ speedup, confirming that merely reducing kernel launches
is futile without eliminating underlying GM bottlenecks.

\textit{Compared with FlashQLA:} Although FlashQLA outperforms FLA via
partial operator fusion and segment-level parallelism, its inter-kernel
execution still materializes intermediate tensors, leaving data locality
fundamentally unresolved. Tailored strictly for memory-bound edge NPUs,
\sysname directly targets these pain points via full fusion for data
locality, operator reordering for buffer efficiency, and layout planning
for minimal transform overheads. These memory-centric optimizations
yield direct hardware benefits: compared with FlashQLA, \sysname reduces
GM traffic by $21.2\%$--$40.8\%$ and L1-to-L0 transfers by
$39.2\%$--$58.4\%$ (\Sref{sec:evaluation-memory-traffic-utilization}), which
unblocks the compute unit to boost Matrix-unit utilization by up to
$1.74\times$ (\Sref{sec:evaluation-memory-traffic-utilization}), ultimately
translating into a $1.79$--$2.41\times$ latency speedup for 4,096-token inputs.
\par
\endgroup

\para{Energy Consumption.}
On Ascend 310P for 4,096-token sequences, \sysname achieves energy reductions of $11.92$--$18.50\times$ over vLLM-Ascend, $4.48$--$4.95\times$ over FLA, $4.47$--$5.01\times$ over FLA-MegaKernel, and $1.76$--$2.11\times$ over FlashQLA. 
By confining states on-core, maximizing tensor reuse, and avoiding unnecessary layout transforms, \sysname reduces the energy-intensive GM traffic that dominates baselines.

\subsection{End-to-End Performance}

\sysname targets only the LA
prefill stage, relying on the native backend for all decoding. Our evaluated
workloads include TriviaQA (9,813--13,717 input vs.\ 7--9 output tokens),
HotpotQA (11,003--13,407 input vs.\ 5--9 output tokens), MMLU-Pro
(2,099--2,267 input vs.\ 1--2 output tokens), and the multimodal MMMU
(2,138--5,621 input vs.\ 2--7 output tokens). 
Specifically, for Qwen3.5-2B on HotpotQA via vLLM-Ascend, prefill consumes 97.1\% of total latency (5.983\,s prefill vs.\ 0.163\,s decode), making it end-to-end bottleneck here.

\para{End-to-end Latency.}
\textbf{\sysname consistently achieves the lowest latency across all 18 model-dataset configurations (Figure~\ref{fig:e2e-dataset-latency}).} Compared with vLLM-Ascend, FLA, FLA-MegaKernel, and FlashQLA, it delivers speedups of 1.60–2.03x, 1.22–1.39x, 1.22–1.36x, and 1.06–1.12x across the Qwen3.5 series. This superiority extends to KDA-based models: yielding up to 2.03x and 1.27x speedups over vLLM-Ascend and FlashQLA for AFM-4.5B, and up to 1.84x and 1.21x for MoE-augmented Ling-3.0-tiny.
With identical peripheral operators and decoding backends across baselines, these system-level speedups validate \sysname's core optimizations.

\para{End-to-End Energy.} System-wide, \sysname reduces total device energy per request by $32.8$--$48.0\%$, $11.8$--$24.1\%$, $11.6$--$23.1\%$, and $6.3$--$17.1\%$ against the four respective baselines. 
Thus, \sysname's kernel-level GM optimizations directly translate into on-device energy savings.

\subsection{Accuracy}
\label{sec:evaluation-accuracy}
Evaluated against a CPU FP64 reference across 360 GDN instances, \sysname achieves maximum NMSEs of $5.23 \times 10^{-8}$ for outputs and $9.38 \times 10^{-12}$ for final states, corresponding strictly to inherent hardware rounding limits.
Further, \sysname preserves end-to-end model quality, yielding negligible F1 score deviations ($\le 0.25$) on both HotpotQA and TriviaQA compared with NPU-native vLLM-Ascend (Table~\ref{tab:gdn-model-quality}).

\begin{table}[!t]
    \centering
    \captionsetup{skip=7pt}
    \caption{Model output quality on Ascend 310P (F1 $\uparrow$).}
    \label{tab:gdn-model-quality}
    \small
    \setlength{\tabcolsep}{2.5pt}
    \setlength{\aboverulesep}{1pt}
    \setlength{\belowrulesep}{1.5pt}
    \renewcommand{\arraystretch}{1.0}
    \begin{tabular}{@{}lrr@{\hspace{8pt}}rr@{}}
        \toprule
        & \multicolumn{2}{c}{\textbf{Qwen3.5-2B}}
        & \multicolumn{2}{c}{\textbf{Qwen3.5-4B}} \\
        \cmidrule(lr){2-3}\cmidrule(l){4-5}
        \textbf{Method}
        & HotpotQA
        & TriviaQA
        & HotpotQA
        & TriviaQA \\
        \midrule
        vLLM-Ascend & 48.90 & 83.70 & 65.39 & 89.17 \\
        \textbf{\sysname} & 49.06 & 83.53 & 65.14 & 89.17 \\
        \bottomrule
    \end{tabular}
\end{table}

\begin{figure}[!t]
    \centering
    \sysnamegraphics[trim=0 10pt 0 12pt,clip]{\columnwidth}{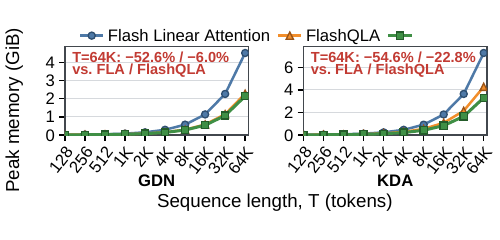}{199.70154}{85.32848}{8}{38.05845}
    \caption{Peak memory consumption of GDN and KDA on Ascend 310P ($H=32$).}
    \label{fig:peak-device-memory}
\end{figure}

\subsection{Peak Device Memory}

Figure~\ref{fig:peak-device-memory} evaluates the kernel-level peak device memory footprint during forward execution. 
By maximizing on-core residency and eliminating transient GM intermediates, \sysname's head-level fusion bounds the peak memory strictly below all baselines.
Specifically, compared with FLA, \sysname reduces peak memory by $31.7$--$52.6\%$ for GDN and $38.7$--$54.6\%$ for KDA. 
Compared with FlashQLA at 65,536 tokens, it saves 6.0\% (GDN) and 22.8\% (KDA).
The larger KDA savings stem from FlashQLA's GDN-centric design: its extra context-parallelism stage materializes a cumulative-gate tensor, which for KDA is $128\times$ larger than for GDN. \sysname naturally avoids this materialization via head-level fusion.

\begin{figure}[!t]
    \centering
    \captionsetup[subfigure]{skip=2pt}
    \sysnamegraphics{\columnwidth}{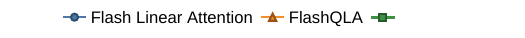}{191.43594}{6.92969}{8}{58.56406}
    \par\vspace{-7pt}
    \begin{subfigure}[t]{\columnwidth}
        \centering
        \includegraphics[width=0.97\linewidth,trim=0 0 0 2pt,clip]{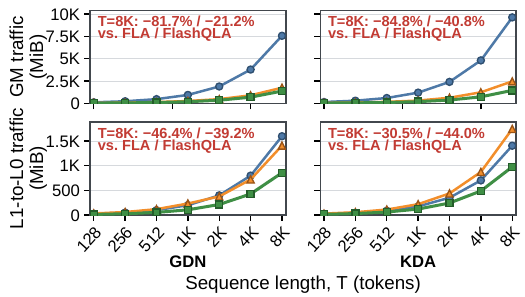}
        \caption{Memory Traffic}
        \label{fig:io-aware-mechanism-memory}
    \end{subfigure}
    \par\vspace{2pt}
    \begin{subfigure}[t]{\columnwidth}
        \centering
        \includegraphics[width=0.97\linewidth]{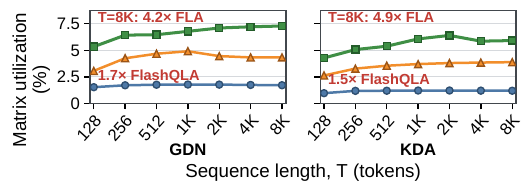}
        \caption{Matrix Utilization}
        \label{fig:io-aware-mechanism-computation}
    \end{subfigure}
    \caption{Memory traffic and Matrix utilization of GDN and KDA on Ascend 310P ($H=32$).}
    \label{fig:io-aware-mechanism}
\end{figure}

\subsection{Memory Traffic and Matrix Utilization}
\label{sec:evaluation-memory-traffic-utilization}

\para{Memory Traffic.} \sysname drastically curtails both global and on-core data movement (Figure~\ref{fig:io-aware-mechanism}(\subref{fig:io-aware-mechanism-memory})). Compared with FLA, it reduces GM traffic by $65.5$--$84.8\%$ and L1-to-L0 transfers by $30.5$--$46.4\%$. Against FlashQLA, it cuts GM traffic by $21.2$--$40.8\%$ and L1-to-L0 transfers by $39.2$--$58.4\%$.
These dual savings stem directly from our co-design: head-level mapping keeps cross-stage intermediates on-core to slash GM traffic, while operator reordering eliminates redundant L1-to-L0 reloads via inter-GEMM operand reuse. The wider GM gap against FlashQLA on KDA reconfirms the structural overhead of its inter-kernel tensor materialization.

\para{Matrix Utilization.}
\sysname eliminates redundant computations through operator rewriting,
reducing Matrix MACs (e.g., by $17.5\%$ and $12.9\%$ versus FLA and
FlashQLA for 4,096-token GDN). Despite executing fewer MACs, it achieves
$5.33$--$7.27\%$ Matrix utilization for GDN and $4.27$--$6.38\%$ for KDA,
outperforming FLA by up to $5.24\times$ and FlashQLA by up to $1.74\times$
(Figure~\ref{fig:io-aware-mechanism}(\subref{fig:io-aware-mechanism-computation})).
This dual advantage indicates that our localized execution and optimized
layouts reduce data-movement overhead and improve Matrix-unit utilization.

\begin{figure}[t]
    \centering
    \includegraphics[width=0.95\columnwidth]{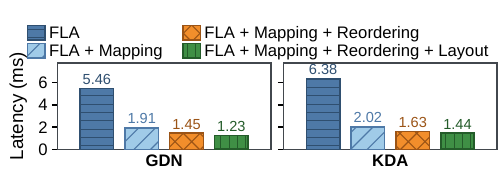}
    \caption{Ablation study of \sysname on Ascend 310P
    ($T=1{,}024$; $H=16$).}
    \label{fig:ablation}
\end{figure}

\subsection{Ablation Study}

We conduct a breakdown analysis of \sysname's techniques (Figure~\ref{fig:ablation}).
Three techniques are represented by Mapping
(\Sref{sec:recurrence-aware-elastic-fusion}), Reordering
(\Sref{sec:state-aware-on-chip-graph-reorganization}), and Layout
(\Sref{sec:interface-layout-planning}).
\textbf{We observe that all three techniques contribute significantly and cumulatively to overall performance.}
Specifically, starting from FLA, we progressively enable each
optimization: First, \textit{head-level mapping} yields the largest
initial speedups ($2.86\times$ for GDN and $3.15\times$ for KDA)
by minimizing cross-operator global memory transfers
while utilizing compute cores.
Next, \textit{lifetime-aware operator reordering} further cuts latency
by 23.8\% and 19.7\% by shortening tensor lifetimes to maximize on-core reuse.
Finally, \textit{dataflow layout planning} trims the remaining
latency by 15.1\% and 11.2\% by balancing layout efficiency against
transformation overheads.

    \section{Related Work}
\label{sec:related-work}

\para{LLM Inference on Edge NPUs.}
Leveraging NPU efficiency \cite{chen2014diannao,chen2014dadiannao,jouppi2017tpu,jouppi2023tpuv4},
recent studies actively optimize on-device LLM/\allowbreak{}VLM inference~\cite{llm-npu,xue2024powerinfer2,chen2025heteroinfer,yin2026shadownpu,hao2026mobile-npu,wei2025tman,zhang2026vlmcache,chen2025sdnpu,lu2025bluelmv,lu2025genieblue,wei2026agentxpu,mllm,executorch,litert-lm,powerserve}.
Key optimizations include \textit{llm.npu}'s fixed-size chunking for static
graphs~\cite{llm-npu}, \textit{PowerInfer-2}'s CPU-NPU pipelining to overlap
weight loading~\cite{xue2024powerinfer2}, \textit{ShadowNPU}'s sparsity-guided
token selection~\cite{yin2026shadownpu}, and \textit{T-MAN}'s table-lookup
low-bit inference~\cite{wei2025tman}.
While these works target standard dense or sparse architectures, \sysname
bridges this gap by co-designing the emerging LA dataflow
specifically for edge NPUs.

\para{Linear Attention in LLMs.}
Recent works optimize linear attention's memory mechanisms, boosting accuracy while retaining linear complexity through fixed-size recurrent states \cite{katharopoulos2020linear,schlag2021fastweights,retnet,gla,arora2024based,aksenov2024rebased,qin2024hgrn2,deltanet,zhang2024gsa,gated-deltanet,kimi-linear,hatamizadeh2026gdn2}.
Specifically, while \textit{RetNet} and \textit{GLA} introduce temporal decay
to control memory retention~\cite{retnet,gla},
\textit{DeltaNet}, \textit{GDN}, and \textit{KDA} enhance state updates through
error-correcting delta rules and fine-grained
gating~\cite{schlag2021fastweights,deltanet,gated-deltanet,kimi-linear}.
These LA variants, particularly GDN and KDA, underpin practical hybrid LLMs
such as \textit{Qwen3.5}~\cite{qwen35-model-card},
\textit{Kimi K3}~\cite{kimi-k3}, and
\textit{GLM-5.3-Flash}~\cite{glm53-flash-kda}.

\para{Linear Attention on GPU/NPU.}
Recent works make LA practical via optimized kernels and runtime
support~\cite{qin2024lightning,chen2026metaattention,beck2025tfla}.
On GPUs, frameworks like \textit{llama.cpp}~\cite{llamacpp-gdn},
\textit{FLA}~\cite{gla,yang2024fla}, and
\textit{FlashQLA}~\cite{flashqla2026} accelerate LA via chunkwise parallelism,
enhanced concurrency, and warp specialization.
For NPUs, basic execution paths in \textit{vLLM-Ascend}~\cite{vllm-ascend}
and \textit{SGLang}~\cite{sglang-linear-attention} or ported GPU designs
fail to exploit memory-constrained edge devices.
\sysname unlocks this potential by co-designing LA execution with edge NPU
hardware.

    \section{Conclusion}
\label{sec:conclusion}

This paper presents \sysname, a memory-efficient LA inference
system for edge NPUs that reduces prefill latency and energy consumption.
\sysname integrates head-level mapping,
lifetime-aware operator reordering, and layout planning to
improve data locality.
Extensive experiments on GDN and KDA demonstrate its effectiveness,
achieving up to $2.03\times$ end-to-end speedup and $48.0\%$ lower device
energy.

\end{document}